\documentstyle[twoside,fleqn,espcrc2,epsf]{article}

\newcommand{\AmS}{{\protect\the\textfont2
  A\kern-.1667em\lower.5ex\hbox{M}\kern-.125emS}}

\newcommand{\be}{\begin{equation}}
\newcommand{\ee}{\end{equation}}
\newcommand{\ben}{\begin{eqnarray}}
\newcommand{\een}{\end{eqnarray}}

\title{Scaling Analysis of Chiral Phase Transition for Two Flavors of
Kogut-Susskind Quarks\thanks{presented by M. Okawa}}

\author{JLQCD Collaboration\\[2mm]
	S. Aoki\address{Institute of Physics, University of Tsukuba,
        Tsukuba, Ibaraki 305, Japan},
        M. Fukugita\address{Institute for Cosmic Ray Research,
        University of Tokyo, Tanashi, Tokyo 188, Japan},
        S. Hashimoto\address{Computing Research Center,
        High Energy Accelerator Research Organization (KEK),\\
        Tsukuba, Ibaraki 305, Japan},
        N. Ishizuka$^{\rm a}$,
	Y. Iwasaki$^{\rm a,d}$,
	K. Kanaya$^{\rm a,}$\address{Center for Computational Physics,
        University of Tsukuba, Tsukuba, Ibaraki 305, Japan},
	Y. Kuramashi\address{Institute of Particle and Nuclear Studies,
        High Energy Accelerator Research Organization (KEK),
        Tsukuba, Ibaraki 305, Japan},\\
        H. Mino\address{Faculty of Engineering, Yamanashi University,
        Kofu 400, Japan},
	M. Okawa$^{\rm e}$,
	A. Ukawa$^{\rm a}$,
	T. Yoshi\'e$^{\rm a,d}$
}

\begin{document}

\begin{abstract}

Report is made of a systematic scaling study of
the finite-temperature chiral phase
transition of two-flavor QCD with the Kogut-Susskind quark action
based on simulations on $L^3\times4$ ($L$=8, 12 and 16) lattices at
the quark mass of $m_q=0.075, 0.0375, 0.02$ and 0.01.
Our finite-size data show that a phase transition is absent for
$m_q\geq 0.02$, and quite likely also at $m_q=0.01$.
The scaling behavior of susceptibilities as a function of $m_q$
is consistent with a second-order transition at $m_q=0$.
However, the exponents deviate from
the $O(2)$ or $O(4)$ values theoretically expected.

\end{abstract}

\maketitle

\section{Introduction}

Study of full QCD thermodynamics with the Kogut-Susskind quark action has
been pursued over a number of years.
A  basic question for this system is the
order of chiral phase transition for light quarks.  For the case of two
flavors, this question was examined by finite-size scaling studies
carried out around
1989-1990\cite{founf2,columbianf2}.
On lattices with the temporal size $N_t=4$ and the quark mass in the
range $m_q=0.025-0.01$,
it was found that the peak height of
susceptibilities increases up to a spatial lattice size $L=12$,
but stays constant within errors between $L=12$ and 16.
The conclusion then was that a phase transition is
absent down to $m_q \approx$ 0.01, which was thought consistent with the
transition being of second-order at $m_q=0$ as suggested by the sigma model
analysis\cite{sigmamodel}.

A more detailed study based on universality argument was recently
attempted\cite{karsch,karschlaermann}.
Critical exponents were extracted from the quark mass dependence of the
critical coupling and the peak height of various susceptibilities
on an $8^3 \times 4$ lattice with $m_q$=0.075, 0.0375 and 0.02.
It was found that the magnetic exponent is
in reasonable agreement with that of the $O(4)$ spin model expected
from universality arguments\cite{sigmamodel}, while the thermal
exponent shows a sizable deviation from the $O(4)$ value.

We have attempted to systematically extend the previous studies both
regarding the spatial volume dependence
and the quark mass dependence to further examine the universality nature
of the transition. For this purpose
we have carried out simulations on lattices of spatial size $L=8, 12$ and $16$
at the quark mass of $m_q=0.075, 0.0375, 0.02$ and $0.01$ in lattice units.
In this article we report on
results of scaling analyses based on these runs\cite{preliminary}.
Studies similar to ours are being carried out by other
groups\cite{laermann,toussaint}.

\section{Simulation}

The full QCD system we study is defined by the partition function
\begin{equation}
Z=\int\prod {\rm d} U_l \exp(S_g)\det(D)^{N_f/4}
\end{equation}
with $S_g$ the standard single-plaquette gauge action, and $D$ the
Kogut-Susskind quark operator.  Simulations are made
on $L^3 \times 4$ lattices with $L$ = 8, 12 and 16.  For the quark mass
$m_q$, we employ $m_q$ = 0.075, 0.0375, 0.02 and 0.01 for each spatial lattice
size $L$.
The hybrid $R$ algorithm\cite{hybridR} is adopted to update gauge
configurations.  In Table~\ref{tab:runs},
we list the values of $\beta$ where our runs are
made.  To control systematic errors of the algorithm,
we choose the molecular dynamics step size to be $\delta\tau\approx m_q/2$
as listed in Table~\ref{tab:runs}.
For each run, 10000 trajectories of unit length are generated starting
from an ordered configuration.  Two runs are made for
$m_q=0.01$ on a $12^3\times 4$ lattice since the first run at $\beta=5.266$
appears to be predominantly in the low-temperature phase
(see Fig.~\ref{fig:histories} below).
Critical exponents we obtain for $L=12$ using two runs separately,
however, agree within our statistical errors.
We therefore show results obtained with the first run in this article.

Inversion of the quark operator
is made with the conjugate gradient algorithm,
reducing the number of floating point operations
by half through the even-odd decimation procedure.  The stopping condition
for the even part of the source vector $b_e$ is
$\sqrt{{\vert\vert b_e-(D^\dagger Dx)_e\vert\vert^2}/3V} < 10^{-6}$
with $V$ the space-time volume
$V=L^3\times 4$.

Observables are calculated at every trajectory.
For computing average values of observables we discard the initial 2000
trajectories of each run.  The errors are estimated by the Jackknife method
with a bin size of 800 trajectories.
Values of observables in the region of $\beta$ around the
simulation point are evaluated by the standard reweighting
technique\cite{reweighting}.

The numerical calculations have been performed on the Fujitsu VPP500/80
supercomputer at KEK.

\begin{table}
\begin{center}
\caption{Parameters of our runs.}
\label{tab:runs}
\vspace*{2mm}
\begin{tabular}{lllll}
\hline
$L$&$m_q=0.075$&$0.0375$&$0.02$&$0.01$\\
     &$\ \delta\tau=0.05$&0.02&0.01&0.005\\
\hline
8     &$\ \ \beta=5.35$&5.306&5.282&5.266\\
12    &\ \ \ \ \ \ \ \ 5.348&5.306&5.282&5.266\\
      &     &     &     &5.2665\\
16    &\ \ \ \ \ \ \ \ 5.345&5.306&5.282&5.266\\
\hline
\end{tabular}
\end{center}
\vspace*{-0.7cm}
\end{table}

\section{Observables}

In the course of our simulation, we measure the following susceptibilities:
\ben
\chi_m&=&V\left[\langle\left(\overline{\psi}\psi\right)^2\rangle-
\langle\overline{\psi}\psi\rangle^2\right],\\
\chi_{t,f}&=&V\left[\langle \left(\overline{\psi}\psi\right)
                            \left(\overline{\psi}D_0\psi\right)\rangle
             -\langle\overline{\psi}\psi\rangle
                            \langle\overline{\psi}D_0\psi\rangle\right]
                \label{eq:sustebegin}\\
\chi_{t,i}&=&V\left[\langle \left(\overline{\psi}\psi\right)P_i\rangle
             -\langle\overline{\psi}\psi\rangle
                            \langle P_i\rangle\right],\\
\chi_{e,f}&=&V\left[\langle\left(\overline{\psi}D_0\psi\right)^2\rangle-
\langle\overline{\psi}D_0\psi\rangle^2\right],\\
\chi_{e,i}&=&V\left[\langle \left(\overline{\psi}D_0\psi\right)P_i\rangle
             -\langle\overline{\psi}D_0\psi\rangle
                            \langle P_i\rangle\right],\\
\chi_{e,ij}&=&V\left[\langle P_iP_j\rangle
            -\langle P_i\rangle\langle P_j\rangle\right],
\label{eq:susteend}
\een
where $D_0$ denotes the temporal component of the Dirac operator,
$i,j=\sigma, \tau$,
and $P_{\sigma,\tau}$ the spatial and temporal plaquette.

Calculation of the fermionic susceptibilities $\chi_m$, $\chi_{t,f}$
and $\chi_{e,f}$ is non-trivial because of the presence of disconnected
double quark loop contributions.
We use the volume source method without gauge
fixing\cite{kuramashi} to evaluate these susceptibilities.

Let us illustrate our procedure for $\chi_m$.
Performing quark contractions and correcting for the flavor factor
arising from the four-flavor nature of the Kogut-Susskind quark field,
we find
\ben
\chi_m&=&\chi_{disc}+\chi_{conn},\\
\chi_{disc}&=&\left(\frac{N_f}{4}\right)^2\frac{1}{V}\Bigl[
\langle\left(\mbox{Tr}D^{-1}\right)^2\rangle\nonumber\\
&&\qquad\qquad\qquad -\langle\mbox{Tr}D^{-1}\rangle^2
\Bigr],\\
\chi_{conn}&=&-\frac{N_f}{4}\frac{1}{V}\sum_{x,y}\langle
D_{x,y}^{-1}D_{y,x}^{-1}\rangle.
\een
Let us define the quark propagator for unit source placed at every
space-time site with a given color $b$ by
\be
G_x^{a,b}\equiv\sum_y \left(D^{-1}\right)_{x,y}^{a,b}.
\ee
From $G_x^{a,b}$, we calculate four quantities
$O_i (i=1,4)$ defined by
\ben
O_1&=&\sum_{x,y}\sum_{a,b}G_x^{a,a}G_y^{b,b},\\
O_2&=&\sum_{x,y}\sum_{a,b}G_x^{a,b}G_y^{b,a},\\
O_3&=&\sum_{x}\sum_{a,b}G_x^{a,a}G_x^{b,b},\\
O_4&=&\sum_{x}\sum_{a,b}G_x^{a,b}G_x^{b,a}.
\een
It is then straightforward to show that
\ben
\left(\mbox{Tr}D^{-1}\right)^2&=&+\frac{9}{8}O_1-\frac{3}{8}O_2
                        -\frac{1}{8}O_3\nonumber\\
&&\hfill +\frac{3}{8}O_4,\\
\sum_{x,y}D_{x,y}^{-1}D_{y,x}^{-1}&=&-\frac{3}{8}O_1+\frac{9}{8}O_2
                        +\frac{3}{8}O_3\nonumber\\
&&\hfill -\frac{1}{8}O_4,
\een
up to terms which are gauge non-invariant, and hence do not contribute
to the average over gauge configurations.
We note that $O_1$ contains connected contributions in addition to the
dominant disconnected double quark loop contribution,
and {\it vice versa} for $O_2$.
The terms $O_3$ and $O_4$ represent contact contributions in
which the source and sink points of quark coincide.

\section{Finite-size scaling analysis}

\begin{figure}[bt]
\centerline{\epsfxsize=75mm \epsfbox{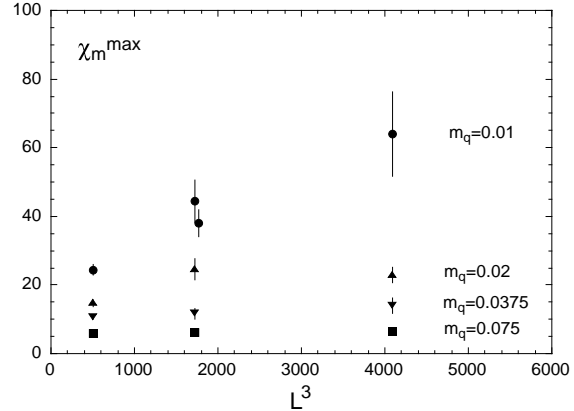}}
\vspace*{-9mm}
\caption{Peak height of the chiral susceptibility $\chi_m$ as a function of
spatial volume $L^3$.  For $L=12$ and $m_q=0.01$ the upper point is from the
run at $\beta=5.266$ and the lower one from $\beta=5.2665$. }
\label{fig:heightvsvolume}
\vspace*{-0.7cm}
\end{figure}

\begin{figure}[bt]
\centerline{\epsfxsize=75mm \epsfbox{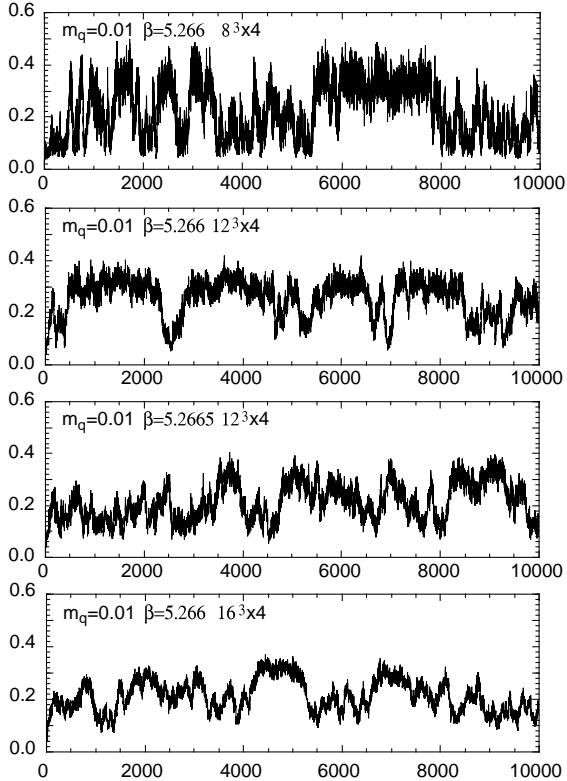}}
\vspace*{-9mm}
\caption{Time history of the chiral order parameter $\overline{\psi}\psi$
for the runs with $m_q$=0.01.}
\label{fig:histories}
\vspace*{-0.7cm}
\end{figure}

We start examination of our data with an analysis of spatial volume dependence
of susceptibilities for each quark mass.
Let $\chi_m^{max}$ be the peak height of $\chi_m$ as a function of $\beta$
evaluated with the reweighting technique.
In Fig.~\ref{fig:heightvsvolume}
we plot the peak height $\chi_m^{max}$ as a function of the spatial volume.

For the heavier quark masses of $m_q=0.075$ and 0.0375 the peak height
increases little over the sizes $L=8-16$, clearly showing that
a phase transition is absent for these masses.  For $m_q=0.02$
an increase of the peak height is seen
between $L=8$ and 12.  The increase, however does not continue beyond
$L=$12; the peak height stays constant within errors between $L=12$ and 16.
We conclude absence of a phase transition also for $m_q=0.02$
confirming the previous work\cite{founf2,columbianf2}.

For the lightest quark mass $m_q=0.01$  employed in our simulation,
we observe a large increase of the peak height
between $L=8$ and 12.  Furthermore, the increase continues up to $L=16$.
The size dependence is consistent with a linear behavior in spatial volume,
which one expects for a first-order phase transition. Other susceptibilities
exhibit a similar size dependence as the quark mass is decreased from
$m_q=0.075$ to 0.01.

This behavior contrasts with the results of a previous study\cite{columbianf2}
which found that the peak height of susceptibilities for $L=16$
stays consistent with those for $L=12$ at
$m_q\approx 0.01$\cite{founf2}.
It is likely that a smaller statistics (2500 trajectories\cite{columbianf2}
as compared to 10000 employed here)
led to an underestimate of susceptibilities in ref.~\cite{columbianf2}.

An important question is whether a linear increase seen in
Fig.~\ref{fig:heightvsvolume} could be regarded as evidence for a first-order
phase transition at $m_q=0.01$.  We think that this is not so for several
reasons.  Looking at the time histories
of the chiral order parameter $\overline{\psi}\psi$ shown in
Fig.~\ref{fig:histories},
we observe an apparent flip-flop behavior between two different values of
$\overline{\psi}\psi$ for $L=8$.
However, the time histories for $L=12$ and 16
are more dominated by irregular patterns, and the width of fluctuation
is smaller.  These features are also reflected in the histograms.
While we clearly see a double-peak distribution for $L=8$, it is less
evident for $L=12$ and barely visible for $L=16$.
Furthermore, the width of the
distribution is smaller for larger lattice sizes and the distance
between the position of two possible peaks is narrower.

These observations suggest the possibility that the increase of
the peak height seen for $m_q=0.01$ up to $L=16$ is a transient
phenomenon due to insufficient spatial volume, similar to an increase
observed between $L=8$ and $12$ for $m_q=0.02$.
In order to check this point, we attempt to normalize the lattice volume by
a relevant length
scale, which we take to be the pion correlation length
$\xi_\pi=1/m_\pi$ at zero temperature.
Using a parametrization of available data for pion
mass as a function of $\beta$ and $m_q$ by the MILC
Collaboration\cite{milcthermo}, we
find $\xi_\pi\approx 3.0$ for $m_q=0.02$ and $\xi_\pi\approx 4.4$ for
$m_q=0.01$.  Hence the size $L=8$ for $m_q=0.02$ roughly corresponds to
$L=12$ for $m_q=0.01$, and $L=12$ to $L=16$.  When compared in this
correspondence the histograms for $m_q=0.02$ and 0.01 are similar in shape.
It is quite possible that
the peak height for $m_q=0.01$ levels off if measured on a larger lattice,
{\it e.g.,} $L=24$.

While a definitive conclusion has to await simulations on larger spatial
sizes, we think it likely that a first-order phase transition is
absent also at $m_q=0.01$.

\section{Analysis of quark mass dependence}

\subsection{Scaling laws and exponents}
\label{sec:scaling}

We have seen in the previous section that the spatial volume
dependence of our data do not show
clear evidence of a phase transition down to $m_q=0.01$.  In the
present section we assume that the two-flavor chiral transition is of
second-order occurring at $m_q=0$.  Various scaling laws follow from
this assumption for the quark mass dependence of the susceptibilities,
from which we can extract information about critical exponents.

For a given quark mass $m_q$, let $g_c^{-2}(m_q)$ be the peak position of
the chiral susceptibility $\chi_m$  as a function of the coupling constant
$g^{-2}$ and let $\chi_m^{max}(m_q)$ be the peak height.
These quantities are expected to scale toward $m_q=0$ as
\ben
g_c^{-2}(m_q)&=&g_c^{-2}(0)+c_gm_q^{z_g} \label{eq:zg} \\
\chi_m^{max}(m_q)&=&c_mm_q^{-z_m}. \label{eq:zm}
\een
The peak height of other susceptibilities similarly  scales as
\ben
\chi_{t,i}^{max}(m_q)&=&c_{t,i}\ m_q^{-z_{t,i}},\qquad i=f,\sigma,\tau\\
\chi_{e,i}^{max}(m_q)&=&c_{e,i}\ m_q^{-z_{e,i}},\qquad i=f,\sigma,\tau\\
\chi_{e,ij}^{max}(m_q)&=&c_{e,ij}\ m_q^{-z_{e,ij}},\qquad i,j=\sigma,\tau
\label{eq:zte}
\een
We note that $\chi_{t,i}$ form three singular parts of the thermal
susceptibility
$\chi_t$=$V\left[\langle \left(\overline{\psi}\psi\right)\epsilon\rangle
             -\langle\overline{\psi}\psi\rangle \langle \epsilon\rangle\right]$
with $\epsilon$ the energy density, and $\chi_{e,i}$ and $\chi_{e,ij}$ form
six singular parts of the specific heat
$C=V\left[\langle\epsilon^2\rangle-\langle\epsilon\rangle^2\right]$.
The leading exponents $z_t$ and $z_e$
for $\chi_t$ and $C$ are then given by
$z_t=\mbox{Max}\{z_{t,i}\}$ and
$z_e=\mbox{Max}\{z_{e,i},z_{e,ij}\}$.

\begin{table}[t]
\begin{center}
\setlength{\tabcolsep}{0.2pc}
\caption{Critical exponents extracted by fits of critical coupling and
peak height of susceptibilities for fixed spatial size $L$ as compared to
$O(2), O(4)$\protect\cite{baker,guillou,kanaya} and mean-field (MF) values. }
\label{tab:exponents}
\vspace*{2mm}
\begin{tabular}{lllllll}
\hline
	&$O(2)$	&$O(4)$	&MF			&$L=8$	&$L=12$	&$L=16$\\
\hline
$z_g$	&0.60 	&0.54	&2/3			&0.70(11)&0.74(6)&0.64(5)\\
\hline
$z_m$	&0.79	&0.79	&2/3			&0.70(4)&0.99(8)&1.03(9)\\
\hline
$z_t$	&0.39	&0.33	&1/3		\\
$z_{t,f}$		&	&	&	&0.42(5)&0.75(9)&0.78(10)\\
$z_{t,\sigma}$		&	&	&	&0.47(5)&0.81(10) &0.82(12)\\
$z_{t,\tau}$		&	&	&	&0.47(5)&0.81(9) &0.83(12)\\
\hline
$z_e$	&-0.01	&-0.13	&0		\\
$z_{e,f}$		&	&	&	&0.21(4)&0.28(7)&0.38(7)\\
$z_{e,\sigma}$		&	&	&	&0.25(6)&0.56(11) &0.58(13)\\
$z_{e,\tau}$		&	&	&	&0.22(6)&0.52(10) &0.55(12)\\
$z_{e,\sigma\sigma}$	&	&	&	&0.18(5)&0.46(8) &0.43(10)\\
$z_{e,\sigma\tau}$	&	&	&	&0.20(5)&0.51(9) &0.50(12)\\
$z_{e,\tau\tau}$	&	&	&	&0.19(5)&0.48(9) &0.47(11)\\
\hline
\end{tabular}
\end{center}
\vspace*{-0.7cm}
\end{table}

For a second-order chiral phase transition, we expect the exponents to be
expressed in terms of the thermal and magnetic exponents $y_t$ and $y_m$;
\ben
z_g&=&y_t/y_h, \\
z_m&=&2-d/y_h, \\
z_t&=&1+y_t/y_h-d/y_h, \\
z_e&=&2y_t/y_h-d/y_h.
\een
Therefore two relations exist among the four exponents $z_g, z_m, z_t$ and
$z_e$, which we take to be
\ben
z_g+z_m&=&z_t+1 \label{eq:consistency1}\\
2z_t-z_m&=&z_e.
\label{eq:consistency2}
\een

The natural values to expect for the exponents are those of $O(2)$
corresponding to exact $U(1)$  symmetry of the Kogut-Susskind quark
action at finite lattice spacing.  Sufficiently close to the continuum
limit, we may also expect the $O(4)$ values as predicted by the effective
sigma model analysis.
The possibility of mean-field exponents arbitrarily close to the critical
point has also been suggested\cite{kocickogut}.

\subsection{Results for exponents}

Our results for the exponents are tabulated in Table~\ref{tab:exponents}.
The exponent $z_g$ that governs the
scaling behavior of the critical coupling $g_c^{-2}(m_q)$ is extracted
from the fit of form (\ref{eq:zg}).
We observe that $z_g$  does not have a clear size dependence within our
error of
about 10\%, and that the values are similar to $O(2)$, $O(4)$ or
mean-field predictions, also listed in the Table, within one to two standard
deviations.

\begin{figure}[bt]
\centerline{\epsfxsize=75mm \epsfbox{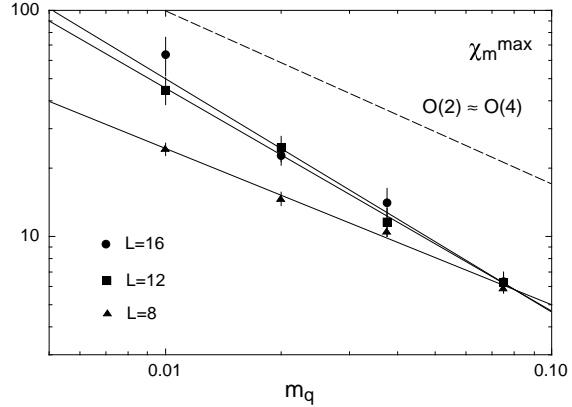}}
\vspace*{-9mm}
\caption{Peak height of the chiral susceptibility $\chi_m$ as a function
of $m_q$ for fixed spatial size $L$.  Solid lines are fits to a single power
(\protect\ref{eq:zm}). Dashed line indicates the slope expected for
$O(2)$ and $O(4)$ exponents which are very similar.}
\label{fig:peakheight}
\vspace*{-0.7cm}
\end{figure}

Let us turn to the exponents determined from the peak height of
susceptibilities.  The values in Table~~\ref{tab:exponents} are
extracted by fits employing a scaling behavior  with a single
power as given in (\ref{eq:zm}--\ref{eq:zte}).
In Fig.~\ref{fig:peakheight} we illustrate the fit for the quark mass
dependence of the peak height of the chiral susceptibility $\chi_m$.

We observe in Table~\ref{tab:exponents}
that all the exponents $z_m, z_t$ and $z_e$
increase as we increase the spatial lattice size $L$.
The value of $z_m$ for the smallest size $L$ =8 is not so different from
the $O(2)$ and $O(4)$ values.  It deviates from the theoretical
prediction for $L$ = 12 and 16, however, and takes a value about 20 \%
larger, which amounts to a two standard deviation difference.

For $z_t$ and $z_e$ various susceptibilities defined
in (\ref{eq:sustebegin})-(\ref{eq:susteend}) generally give consistent
results.
We observe, however,  a $10-20$\% larger value of $z_t$ compared
with the theoretical prediction already for $L=8$, and the discrepancy
increases to a factor two difference for $L=12$ and 16.
The disagreement is more apparent for the exponent $z_e$ for which
values in the range $z_e\approx 0.5-0.6$ are obtained for larger sizes
in contrast to a negative value for the $O(2)$ and $O(4)$ theories.

\begin{figure}[bt]
\centerline{\epsfxsize=75mm \epsfbox{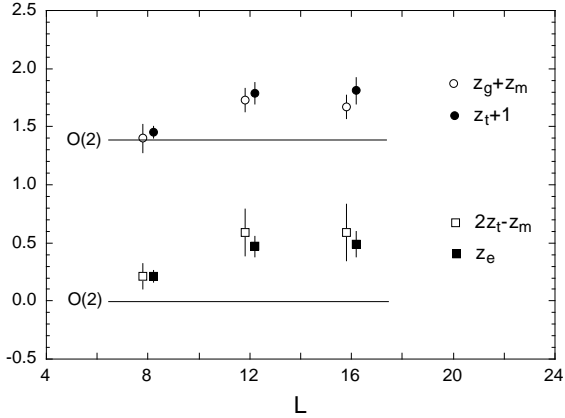}}
\vspace*{-9mm}
\caption{Consistency check of exponents for a given spatial size $L$;
$z_t+1$ against $z_g+z_m$, and  $z_e$ against $2z_t-z_m$. Lines are predictions
for $O(2)$ symmetry.  Values for $O(4)$ are similar.}
\label{fig:consistency}
\vspace*{-0.7cm}
\end{figure}

We have noted in Sec.~\ref{sec:scaling} that the four exponents $z_g, z_m, z_t$
and $z_e$ should satisfy two consistency equations
reflecting the fact
that two relevant operators govern a second-order phase transition.
In Fig.~\ref{fig:consistency}  we plot the two sides of the equations
(\ref{eq:consistency1}) and (\ref{eq:consistency2}) using the
values of exponents given in Table~\ref{tab:exponents}.
For $z_t$ and $z_e$ we take an average over operator combinations since
the values are mutually in agreement within the error.
We observe that the consistency is well satisfied
for each spatial volume even though values of individual exponents deviate from
those of $O(2), O(4)$ or mean-field theory predictions.

We have also attempted fits allowing for a constant term in the fitting
function $\chi_i^{max}=c_{0i}+c_{1i}m_q^{-z_i}$.  We are not able to
obtain reliable fits taking $z_i$ as a free parameter,
since the errors of fitted values are too large.  Fixing  the exponent
$z_i$ to the theoretical $O(2)$ or $O(4)$ values, we find that
the quality of fit
generally worsens compared with the single power fit.
In particular, the fit tends to misses the point for the smallest quark mass
$m_q=0.01$ for $L=16$.

We are led to conclude that the exponents show deviation from
$O(2)$ or $O(4)$ values, at least in the range of quark mass
$m_q=0.075-0.01$ explored in our simulation.

\subsection{Results for scaling function}

\begin{figure}[bt]
\centerline{\epsfxsize=75mm \epsfbox{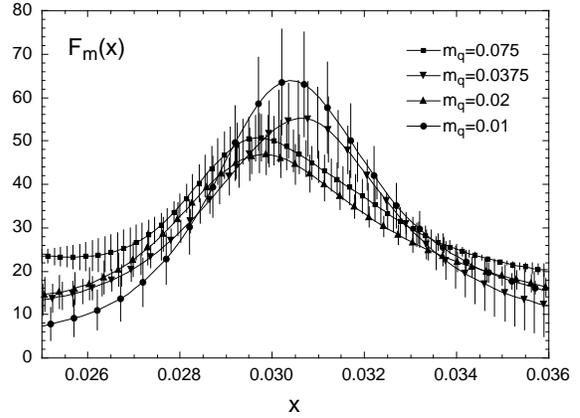}}
\vspace*{-9mm}
\caption{Scaling function $F_m(x)$ normalized as
$\chi_m(g^2,m_q)\cdot (m_q/0.01)^{z_m}$ as a function of
$x=(6/g_c^2(m_q)-6/g_c^2(0))\cdot (m_q/0.01)^{-z_g}$ for $L=16$ with measured
values $z_g=0.6447, z_m=1.033, 6/g_c^2(0)=5.2353$.}
\label{fig:scalingfunction}
\vspace*{-6mm}
\end{figure}

For a second-order phase transition, the singular part of the chiral
susceptibility $\chi_m(g^2,m_q)$ is expected to scale as
\be
\chi_m(g^2,m_q)=m_q^{-z_m}\cdot F_m(x),
\ee
where $F_m(x)$ is a function of scaling variable $x$ which we take to be
\be
x=\left(6/g_c^2(m_q)-6/g_c^2(0)\right)\cdot m_q^{-z_g}.
\ee

We show in Fig.~\ref{fig:scalingfunction} estimates of the scaling function
using data for the size $L=16$.  Both $F_m(x)$ and $x$ are normalized by
the values for $m_q=0.01$, and the measured values are employed for the
exponents: $z_g=0.6447$, $z_m=1.033$ and $6/g_c^2(0)=5.2353$.
Given the magnitude of statistical error which increases from 10\% to 20\%
as $m_q$ decreases from $m_q=0.075$ to $0.01$, we find scaling with respect
to quark mass to be reasonably satisfied.

We have also calculated the scaling function $F_m(x)$ using
the $O(4)$ values for
the exponents\cite{kanaya} $z_g=0.538$, $z_m=0.794$ and the value of
$6/g_c^2(0)$ obtained with a fit of $g_c^2(m_q)$ with the $O(4)$ value for
$z_g$.  We find that scaling worsens.  In particular the curve for the
smallest quark mass $m_q=0.01$ is too high in this case.

\section{Conclusions}

In this article we have reported results of our study of the
two-flavor chiral phase transition with the Kogut-Susskind quark action on an
$N_t=4$ lattice.  Our analysis of the spatial volume dependence of the
peak height of susceptibilities confirms the absence of a phase transition
for $m_q\geq 0.02$ as reported previously\cite{founf2,columbianf2}.
At $m_q=0.01$ the peak height exhibits
an almost linear increase over the sizes
$L=8-16$ contradicting a previous work\cite{columbianf2}.
We have argued, based on an examination of fluctuations of
observables and a consideration of spatial volume normalized by the
zero-temperature pion mass, that the increase is a transient phenomenon
arising from an insufficient spatial volume.
We conclude that a first-order transition is likely to be absent also
at $m_q=0.01$.

We have also found that the quark mass dependence of susceptibilities
is consistent with a second-order transition located at $m_q=0$;
the critical exponents we have obtained satisfy required consistency
conditions, and the susceptibility $\chi_m$ reasonably scales in terms of
variable defined with the measured exponents.
However, the values of exponents
themselves deviate from either $O(2), O(4)$ or mean-field theory predictions.

Further work is needed to elucidate the universality nature of the
two-flavor chiral phase transition in finite-temperature QCD.

\section*{Acknowledgements}

This  work is supported by the Supercomputer Project (No.~1) of High Energy
Accelerator Research Organization (KEK), and also in part by the Grants-in-Aid
of the Ministry of Education (Nos. 08640349, 08640350, 08640404,
08740189, 08740221).

\end{document}